\documentclass[prc,aps,a4paper,groupedaddress,superscriptaddress,nofootinbib
,preprintnumbers,twocolumn]{revtex4}
\usepackage{graphicx}
\usepackage{amsfonts}
\usepackage{amssymb}
\usepackage{amsmath}
\usepackage{natbib}
\usepackage{dcolumn}
\usepackage{bm} 
\usepackage{color,xcolor}
\usepackage{multirow}
\usepackage{caption}
\usepackage{hyperref}
\usepackage[normalem]{ulem}

\newcommand{\bwt}{\begin{widetext}}
	\newcommand{\ewt}{\end{widetext}}
\newcommand{\beq}{\begin{equation}}
	\newcommand{\eeq}{\end{equation}}
\newcommand{\bea}{\begin{eqnarray}}
	\newcommand{\eea}{\end{eqnarray}}

\begin{document}
\title{Strong magnetic fields and pasta phases reexamined}
\author{Luigi Scurto}
\email{lscurto@student.uc.pt}
\affiliation{CFisUC, Department of Physics, University of Coimbra, 3004-516 Coimbra, Portugal}
\author{Helena Pais} 
\email{hpais@usal.es} 
\affiliation{Department of Fundamental Physics, University of Salamanca, E-37008 Salamanca, Spain} 
\author{Francesca Gulminelli}
\email{gulminelli@lpccaen.in2p3.fr}
\affiliation{Normandie Univ., ENSICAEN, UNICAEN, CNRS/IN2P3, LPC Caen, F-14000 Caen, France}

\begin{abstract}

In this work, we compute the structure and composition of the inner crust of a neutron star in the presence of a strong magnetic field, such as can be found in magnetars.
To determine the geometry and characteristics of the crust inhomogeneities, we consider the compressible liquid drop model, where surface and Coulomb terms are included in the variational equations, and we compare our results with previous calculations based on more approximate treatments.
For the equation of state (EoS), we consider two non-linear relativistic mean-field models with different slopes of the symmetry energy, and 
we show that the extension of the inhomogeneous region inside the star core due to the magnetic field strongly depends on the behavior of the symmetry energy in the crustal EoS.
Finally, we argue that the extended spinodal instability observed in previous calculations can be related to the presence of small amplitude density fluctuations in the magnetar outer core, 
rather than to a thicker solid crust. The compressible liquid drop model formalism, while in overall agreement with the previous calculations, leads to a systematic suppression of the metastable solutions, thus allowing a more precise estimation of the crust-core transition density and pressure, and therefore a better estimation of the crustal radius.
\end{abstract}
\maketitle

\section{Introduction}
Neutron star (NS) modeling is a subject of extreme interest in the astrophysical community since the recent emergence of multi-messenger observations. However, their internal structure and composition is still very uncertain and a huge effort is put by the scientific community in order to better understand their properties. Schematically, the interior of a neutron star is divided into four layers, an outer and inner crust and an outer and inner core. In the bottom part of the inner crust region, it is generally accepted that heavy clusters of matter with exotic shapes, called pasta phases, could arise due to the competition between the strong and the electromagnetic interaction \cite{ravenhall83,hashimoto84,horowitz05,watanabe05,maruyama05,avancini08,avancini10,pais12PRL,bao14,Newton22}. From the observational point of view, the static properties of the inner crust, particularly its moment of inertia, should be related to neutron star phenomena such as pulsar glitches \cite{chamel13,link99,andersson12}, while its associated transport properties are important to settle the magneto-thermal evolution of the star \cite{pons2013,Newton2013,Horowitz2015,Lin2020}. 

One category of neutron stars, called magnetars \cite{duncan,thompson,usov,paczynski}, are the source of the strongest magnetic fields observed in nature. These magnetic fields span between $\approx 10^{12}$ G to $\approx 10^{15}$ G on the surface \cite{olausen14,mcgill}, and, according to the scalar virial theorem \cite{lai91,shapiro}, 
may reach even higher values inside the star, up to $\approx 10^{18}$ G, 
a threshold value also predicted in other calculations, where the coupled Maxwell-Einstein equations are solved numerically \cite{lorene1,lorene2,broderick02,chatterjee15,gomes19,cardall01,sengo20}. 

Numerous works have considered how these strong magnetic fields can affect the neutron star outer \cite{Chamel11,Potekhin13,Chamel15,Stein16} and inner \cite{rabhi2,fang16,fang17,chen17,fang17a,pais21,wang22} crust. 
In Ref.~\cite{Mutafchieva19}, the authors study the effects of the magnetic field on both the outer and inner crusts due to the quantization of electron motion. They observed that the inner crust is almost unaffected for fields below $\approx 10^{16}$G, and only marginally affected for higher order fields.

In the case of the inner crust, Thomas-Fermi calculations have been performed at constant proton fraction \cite{nandi,lima}, and more recently in Ref.~\cite{bao} $\beta-$equilibrium matter , that is the appropriate condition when dealing with neutron stars, was considered.
The general result of those studies is that the magnetic field affects the density profiles of the Wigner-Seitz cells, with an increased average proton fraction and an increase of the mass of the clusters in the crust. However the transition densities between the different geometries and the crust-core transition are affected in a very weak and non-monotonous way. In the most recent work, Ref.~\cite{bao}, that used the TM1 and IUFSU functionals, the authors reported a sizable effect only for very strong magnetic fields, $B=10^{18}$ G, for which an important decrease of the core-crust transition density was observed. 

An alternative approach based on a dynamical spinodal calculation was first performed in \cite{fang16,fang17}, where it was shown that new regions of instability arise at densities higher than the one expected for the transition to the homogeneous core, due to the presence of a strong magnetic field. 
In such a calculation \cite{Avancini2005,Providencia2006},  oscillations eigenfrequencies are calculated by considering the matter response to small deviations from equilibrium in the distribution functions of the particles and of the meson fields. In the core region, the frequencies $\omega$ of the modes are real numbers, and the fluctuations are spontaneously damped. Inside the spinodal region, the frequencies are imaginary, reflecting the instability of homogeneous matter with respect to density fluctuations. 
The maximum frequency of the unstable modes $\Gamma=|\omega|$ can be taken as the one that drives the instability, and it measures the growth rate of the density fluctuation in the linear response regime. In Refs.~\cite{fang16,fang17}, only the longitudinal modes that propagate along the magnetic field were considered. Later, in Ref.~\cite{Avancini18}, the authors also calculated the transverse modes, and observed that for magnetic fields of the order of $\approx 10^{18}$G, the spinodal section is reduced, and for fields one order of magnitude lower, the effect is completely washed out, and the transverse spinodal coincides with the $B=0$ field one, meaning that the the propagation of the perturbations perpendicular to the magnetic field are more difficult under the presence of strong magnetic fields, and accounting for longitudinal modes is enough.  In the hypothesis that the linear dynamics is continued until formation of the ion structure that characterizes the solid crust,   
the density and pressure at which the eigenfrequency goes to zero  gives an estimation of the crust-core transition point. In this interpretation, the extended instability zone, due to the effect of the magnetic field, thus corresponds to a higher crust-core transition density, and an extended crust for the neutron star. These results appear in contradiction with the previous findings of Ref.~\cite{bao}, where the authors did not find such a region.

Subsequently, in Refs.~\cite{pais21,wang22}, a coexistence phase (CP) approximation was performed to check the dynamical spinodal results of Refs.~\cite{fang16,fang17}. This calculation was performed both with fixed proton fraction \cite{pais21} and 
at $\beta-$equilibrium \cite{wang22}.
In both Refs.~\cite{pais21,wang22}, the solutions were in agreement with the ones of the instability region found from the dynamical spinodal approach of Refs.~\cite{fang16,fang17}. This points to the fact that the CP approach also predicts extra regions of clusterized matter above the $B=0$ region, and therefore a thicker crust.
However, in \cite{wang22}, it was also shown that these extra regions of clustered matter have different properties with respect to the rest of the inner crust. In particular, very close densities between the clusters and the medium in which they are embedded were found. 
This last finding suggests that the extended crust might be rather interpreted as a slightly inhomogeneous core, with density fluctuations of an amplitude that is too small to be captured by the Thomas-Fermi approach \cite{bao}. A definitive conclusion cannot be reached though, because 
the CP calculation is not a self-consistent approach. Indeed, surface and Coulomb
terms are added after the minimization of the energy density, and therefore the possible influence of the surface properties on the solution of the variational equations is neglected. 

To better settle the issue of the effect of the magnetic field on the crust thickness, in this paper we study the structure of the inner crust of a neutron star in the presence of a strong magnetic field by explicitly including the surface and Coulomb contribution in the variational equations for the inner crust, using the compressible liquid drop (CLD) model  \cite{baym71,lattimer85,lattimer91,bao14,pais15,DinhThi22}.

We also calculate the pasta structures in the CP approximation to compare  with Refs.~\cite{pais21,wang22}, and use the same relativistic mean-field (RMF) functionals as in those previous works, namely the NL3 \cite{lalazissis97} and the NL3$\omega\rho$ \cite{horowitz01,paisVlasov}.

\begin{table*}
    \begin{tabular}{ccccccc}
    \hline
    \hline
          & $B/A$ (MeV) & $\rho_0$ (fm$^{-3}$) & $M^*/M$ & $K$ (MeV) & $\mathcal{E}_{sym}$ (MeV) & $L$ (MeV)  \\
    \hline
        NL3 & 16.24 & 0.148 & 0.60 & 270 & 37.34 & 118 \\
        NL3$\omega\rho$ & 16.24 & 0.148 & 0.60 & 270 & 31.66 & 55\\
    \hline
    \hline
    \end{tabular}
    \caption{Symmetric nuclear matter properties at saturation density for the NL3 \cite{lalazissis97} and NL3$\omega\rho$ \cite{horowitz01,paisVlasov} models. From left to right:  binding energy per baryon, saturation density, normalized nucleon effective mass, incompressibility, symmetry energy and slope of the symmetry energy.}
    \label{Tab1}
\end{table*} 

These two models belong to the same family, i.e. they have the same isoscalar properties. NL3$\omega\rho$ was constructed to model the density dependence of the symmetry energy because NL3 has a very high slope of the symmetry energy at saturation. We should keep in mind though that NL3 should be adequate to study sub-saturation density regimes, like the NS inner-crust, because this model gives a very good description of the properties of stable nuclei. The properties of symmetric nuclear matter at saturation density of these two models can be found in Tab.~\ref{Tab1}.
Both these models predict stars with masses above the $2M_\odot$ \cite{paisVlasov,fortin16}, even when hyperonic degrees of freedom are taken into account, and the NL3$\omega\rho$ model also satisfies the constraints imposed by neutron-matter microscopic calculations \cite{hebeler13,paisVlasov}.

The crust-core transition density strongly depends on the symmetry energy and particularly on its slope at saturation $L$ \cite{xu09,vidana09,ducoin10,ducoin11,newton13,paisVlasov,li20}. This latter quantity is still not yet well constrained. Ab-initio chiral effective field theory calculations seem to favor values for $L$ below 60 MeV \cite{lattimer13}, or below 90 MeV, when astrophysical observations are taken into account as extra constraints \cite{oertel17}. Reed {\it et al.} \cite{Reed21} have performed an analysis on the PREX-2 data \cite{Adhikari21}, obtaining a large value for $L$, $L=106 \pm 37$ MeV. However, other studies, like the one performed by Essick {\it et al.} \cite{Essick:2021kjb}, that predicted $L=53^{+14}_{-15}$ MeV, by also combining astrophysical observations, or the one by Estee {\it et al.} \cite{estee21}, that measured the charged pion spectra at high transverse momenta, suggesting $42<L<117$ MeV, are both compatible with PREX-2 analysis. Moreover, Reinhard {\it et al.} \cite{Reinhard21} based on the PREX-2 results, were able to predict a smaller neutron skin thickness, which lead them to infer a smaller slope of the symmetry energy, $L=54\pm 8$ MeV. Recently, the CREX \cite{CREX:2022kgg} collaboration has measured the $^{48}$Ca neutron skin thickness, and analyses seem to indicate that $L$ could be smaller than PREX-2 predictions. Finally, Mondal {\it et al.} \cite{Mondal2022} have recently shown in a Bayesian analysis that the constraints on $L$ from both PREX-2 and CREX are very loose, if the uncertainties in the surface properties and their correlation with the bulk are accounted for. 

In the present paper, we show how different values of $L$ can lead to substantial differences in the behaviour of the system in the presence of a strong magnetic field.
For both equations of state, the CLD approach is shown to give a very precise estimation of the transition point, because the metastable solutions appearing in the simpler CP formalism of Refs.~\cite{pais21,wang22} are suppressed. 
In qualitative agreement with Ref.~\cite{bao}, the effect of the field on the extension of the crust is seen to be small, even if we do not see the systematic decrease of the transition densities observed in that study. 
We confirm the qualitative effects in the proton density and proton fraction observed in previous works, and additionally show
that the effect of the field is also to induce equilibrium density fluctuations in the outer core. This finding nicely explains the spinodal instabilities observed in the homogeneous matter calculations of Refs.~\cite{fang16,fang17}.

The  paper is organized as follows: in Section II the methods and the formalism are given, in Section III we show  our results and discussion, and finally, in Section IV, conclusions are drawn.

\section{Theoretical Framework}

In this work, NS matter is described within a RMF approximation, where the interaction between the nucleons is mediated by three types of mesons: the isoscalar-scalar meson $\sigma$, the isoscalar-vector meson $\omega$ and the isovector-vector meson $\rho$. In order to achieve electrical neutrality, we also introduce electrons in our description.  
Throughout the work, we consider an electromagnetic field 
of the type $A^\mu=(0,0,Bx,0)$, so that the resulting 
field is oriented along the $z$ axis. 
We take the anomalous magnetic moment of the nucleons to be zero, as it was shown in the previous studies \cite{wang22} that 
its main effect is only to
increase the number of disconnected regions in the spinodal analysis, because of the removal of the spin polarization degeneracy. We use the quantity $B^*$, defined as $B^* = B/B^c_e$, with $B^c_e= 4.414 \times 10^{13}$ G being the critical field at which the electron cyclotron energy is equal to the electron mass.

The Lagrangian density of our system is given by
\begin{equation}
    \mathcal{L}=\sum_{i=p,n}\mathcal{L}_i+\mathcal{L}_e+\mathcal{L}_\sigma+\mathcal{L}_\omega+\mathcal{L}_\rho+\mathcal{L}_{nl}+\mathcal{L}_A \, .
\end{equation}
Here, $\mathcal{L}_e$ and $\mathcal{L}_A$ are the standard electron Lagrangian density and electromagnetic term, given by
\begin{equation}
    \mathcal{L}_e=\bar{\psi}_e\big[\gamma_\mu\big(i\partial^\mu+eA^\mu\big)-m_e\big]\psi_e,
\end{equation}
\begin{equation}
    \mathcal{L}_A=-\frac{1}{4}F_{\mu\nu} F^{\mu\nu} \, ,
\end{equation}
with $F_{\mu\nu}=\partial_\mu A_\nu-\partial_\nu A_\mu$ .
The nucleon Lagrangian density is given by 
\begin{equation}
    \mathcal{L}_i=\bar{\psi}_i\big[\gamma_\mu iD^\mu-M^*\big]\psi_i \, ,
\end{equation}
with
\begin{equation}
    M^*=M-g_\sigma\phi \, ,
\end{equation}
\begin{equation}
    iD^\mu=i\partial^\mu-g_\omega V^\mu-\frac{g_\rho}{2}\mathbf{\tau}\cdot\mathbf{b}^\mu-\frac{1+\tau_3}{2}eA^\mu \, ,
\end{equation}
where $e=\sqrt{4\pi/137}$ is the electron charge, and $\tau_3=\pm 1$ is the isospin projection respectively for protons and neutrons.

The mesonic components of the Lagrangian density are given by 
\begin{equation}
    \mathcal{L}_\sigma=\frac{1}{2}\bigg(\partial_\mu\phi\partial^\mu\phi-m_\sigma^2\phi^2 -\frac{1}{3}\kappa\phi^3-\frac{1}{12}\lambda\phi^4  \bigg) \, ,
\end{equation}
\begin{equation}
    \mathcal{L}_\omega=-\frac{1}{4}\Omega_{\mu\nu}\Omega_{\mu\nu}+\frac{1}{2}m_\omega^2V_\mu V^\mu +\frac{\xi}{4!}\xi g_\omega^4(V_\mu V^\mu)^2 \, ,
\end{equation}
\begin{equation}
    \mathcal{L}_\rho=-\frac{1}{4}\mathbf{B}_{\mu\nu}\cdot\mathbf{B}^{\mu\nu}+\frac{1}{2}m_\rho^2\mathbf{b}_\mu\cdot \mathbf{b}^\mu \, ,
\end{equation}
with the tensors written as
\begin{eqnarray}
\Omega_{\mu\nu}&=&\partial_\mu V_\nu - \partial_\nu V_\mu \, , \\
\mathbf{B}_{\mu\nu}&=&\partial_\mu \mathbf{b}_\nu - \partial_\nu \mathbf{b}_\mu - g_\rho \left(\mathbf{b}_\mu \times \mathbf{b}_\nu \right)\, .
\end{eqnarray}

The NL3$\omega\rho$ model considers an extra term, 
\begin{align}
    \mathcal{L}_{nl}=&\Lambda_{\omega\rho}g_\omega^2g_\rho^2V_\mu V^\mu\mathbf{b}_\mu\cdot \mathbf{b}^\mu \, , \notag
\end{align}
responsible for the density dependence of the symmetry energy.

\begin{figure*}
	\begin{center}
		\begin{tabular}{cc}
			\includegraphics[width=0.45\linewidth,angle=0]{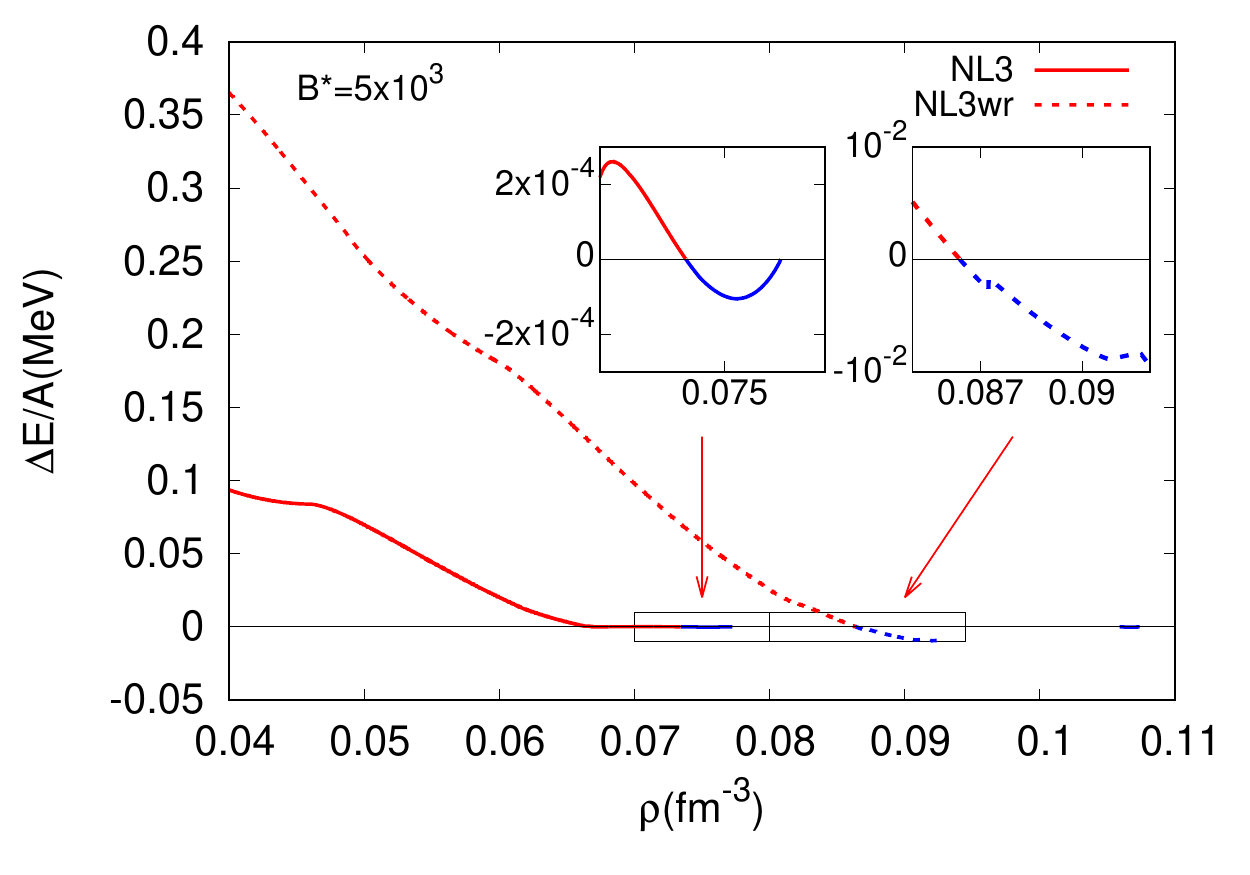}&
			\includegraphics[width=0.45\linewidth,angle=0]{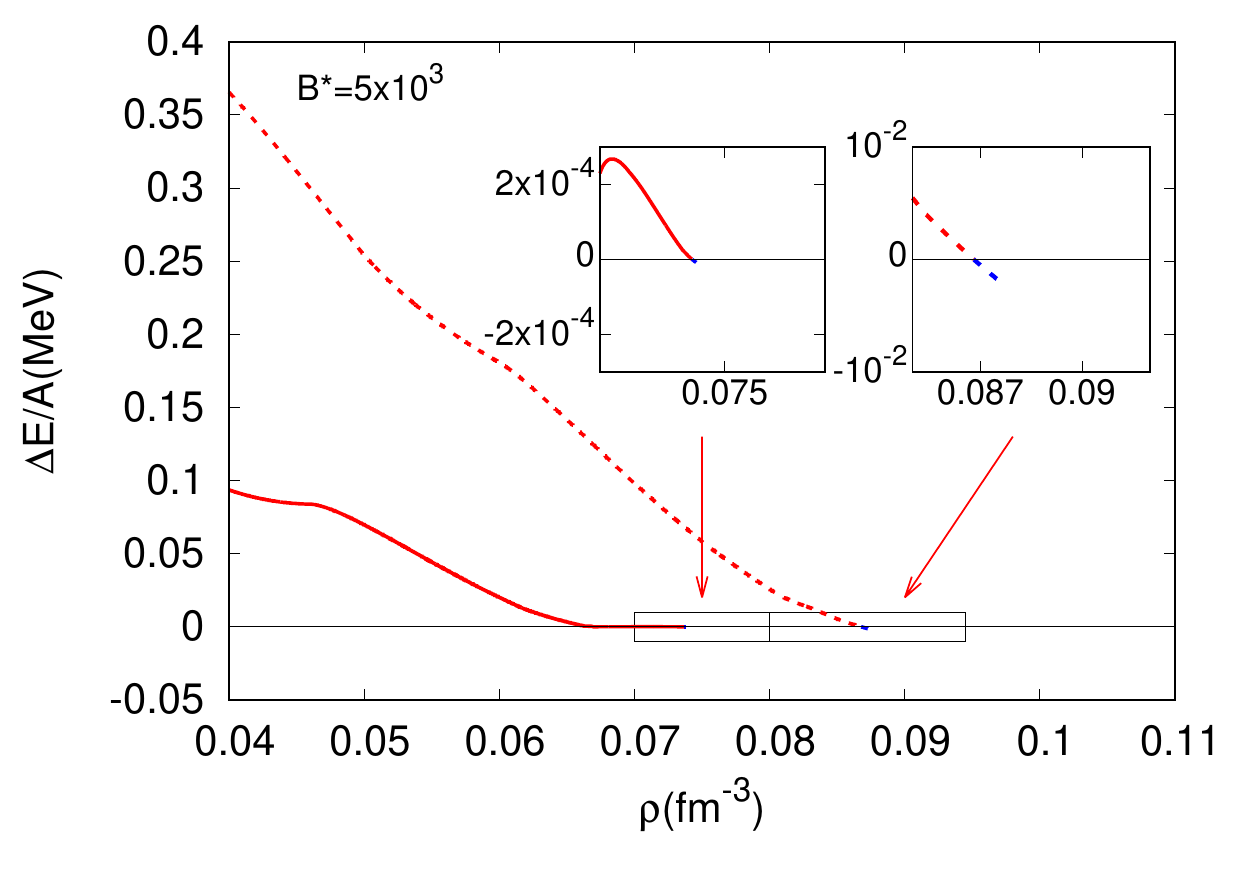}\\
			\includegraphics[width=0.45\linewidth,angle=0]{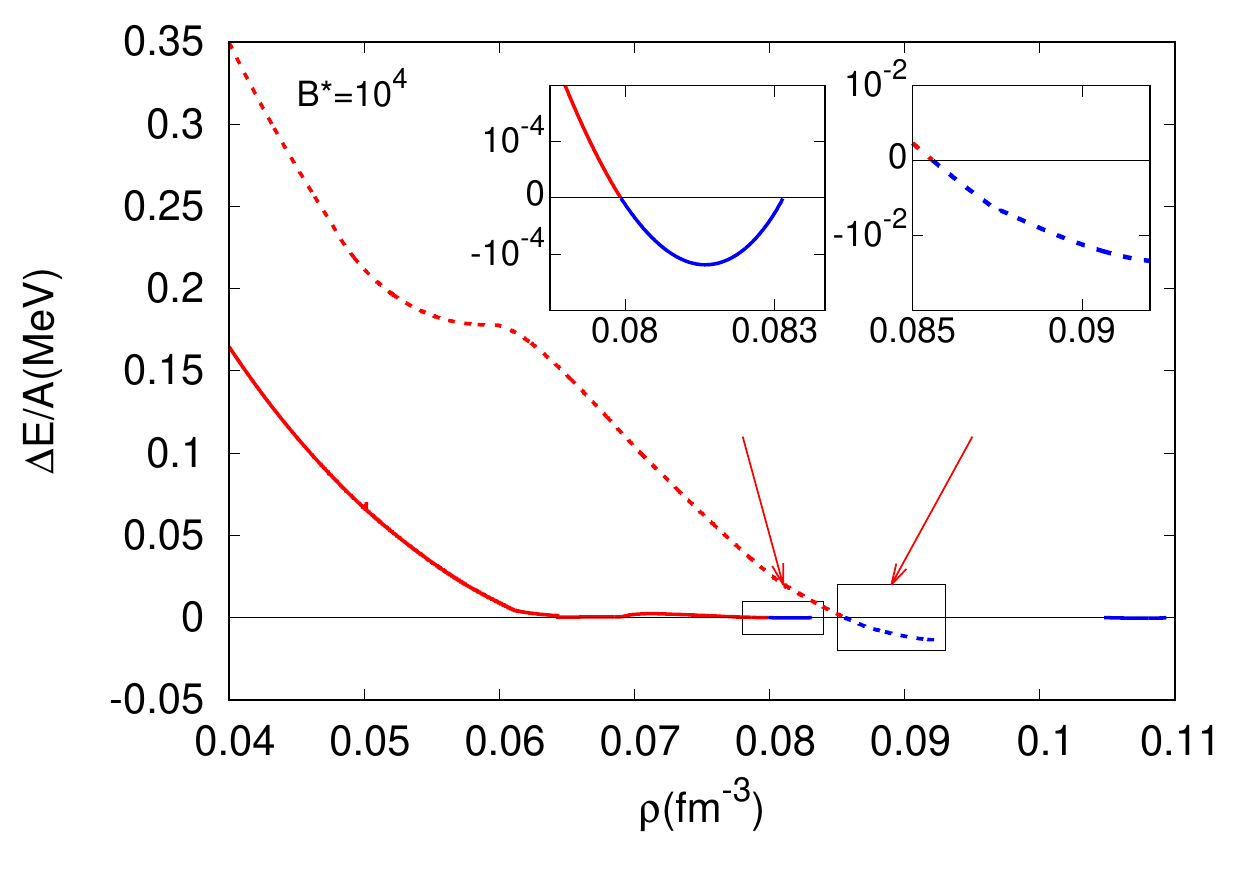} &
			\includegraphics[width=0.45\linewidth,angle=0]{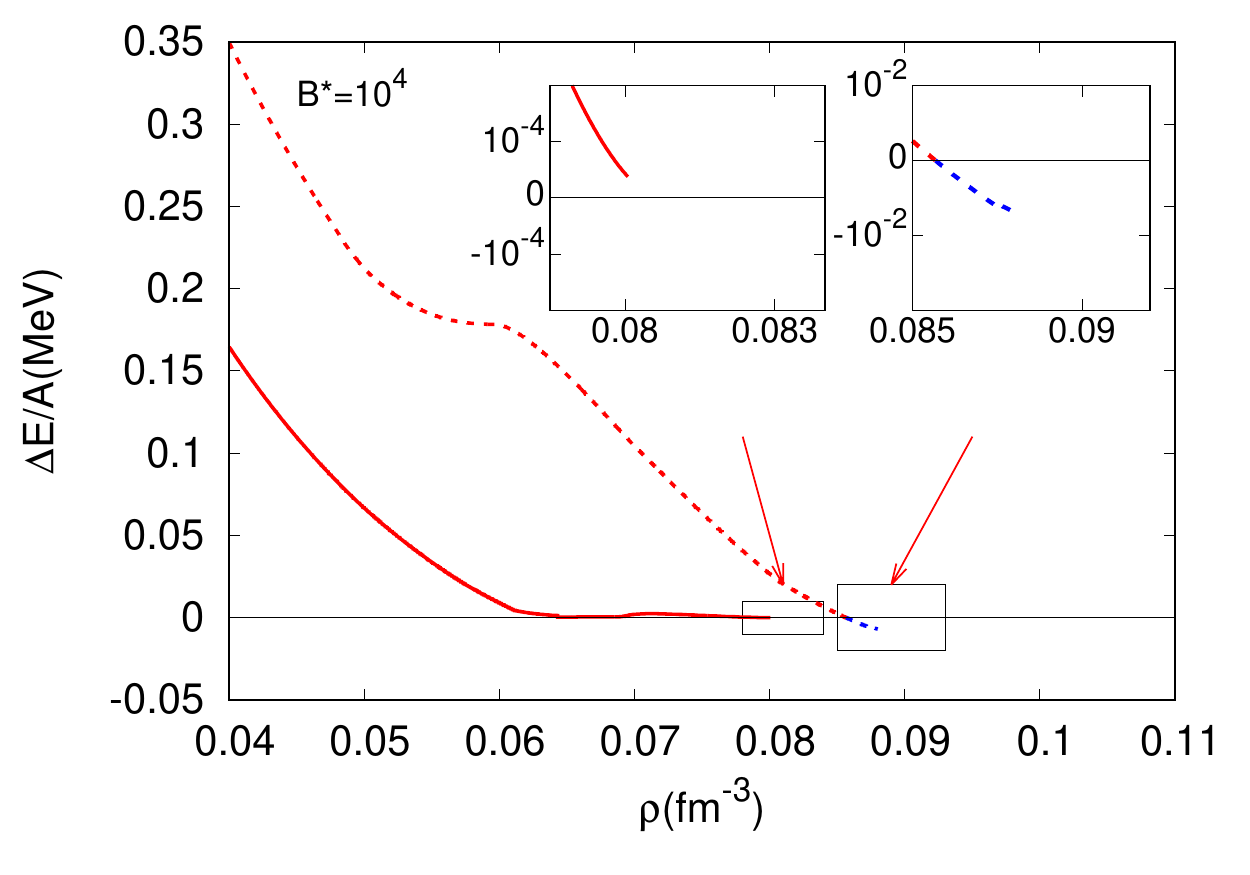}
		\end{tabular}
	\end{center}
	\caption{Difference between the energy per baryon of homogeneous matter and the energy per baryon of clustered matter, as a function of the total baryon density, for the NL3 (solid) and NL3$\omega\rho$ (dashed) models, in a CP (left panels) and CLD (right panels) calculations. The magnetic field intensity is fixed to $B^*=5\times 10^3$ (top panels) and $B^*=10^4$ (bottom panels).  In red, this difference is positive, $E/A_{hm}>E/A_{cl}$, and in blue it is negative. }
	\label{fig1}
 \end{figure*}

From the Euler-Lagrange equations, we get the fields equations of motion in the mean-field approximation. 
As a result, 
the scalar and vector densities for protons and neutrons, and the electron density, are given by

\begin{eqnarray}
    \rho_{s,p}&=&\frac{q_pBM^*}{2\pi^2}\sum_{\nu=0}^{\nu_{\rm max}^p}g_s \ln\bigg|\frac{k^p_{F,\nu}+E^p_F}{\sqrt{M^{*2}+2\nu q_pB}}\bigg| \, , \label{eq:rhosp}
\end{eqnarray}
   
 \begin{eqnarray}
    \rho_{s,n}&=&\frac{M^*}{2\pi^2}\bigg[E^n_Fk^n_{F}-M^{*2}\ln\bigg|\frac{k^n_{F}+E_F^n}{M^*}\bigg|
    \bigg] \, , 
 \end{eqnarray}    

\begin{eqnarray}
    \rho_p&=&\frac{q_pB}{2\pi^2}\sum_{\nu=0}^{\nu_{\rm max}^p} g_s k^p_{F,\nu} \, , \label{eq:rhop} 
\end{eqnarray}

\begin{eqnarray}
    \rho_n&=&\frac{{k^n_{F}}^3}{3\pi^2} \, ,
 \end{eqnarray}

\begin{eqnarray}
    \rho_e&=&\frac{|e|B}{2\pi^2}\sum_{\nu=0}^{\nu_{\rm max}^e} g_s k^e_{F,\nu} \,  , \label{eq:rhoe}
\end{eqnarray}

where $\nu=n+\frac{1}{2}-\frac{1}{2}\frac{q}{|q|}s=0,1,\cdots,\nu_{\rm max}$ enumerates the Landau levels (LL) for fermions with electric charge $q$, $q=e$ for electrons and $q=q_p$ for protons. $s$ is the spin quantum number, $+1$ for spin up cases and $-1$ for spin down cases. The spin degeneracy factor of the Landau levels, $g_s$, is equal to $g_s=1$ for $\nu=0$ and $g_s=2$ for $\nu>0$, and $\nu_{\rm max}$ is the largest LL 
occupied by fully degenerate charged fermions, defined as
\begin{align}
    \nu_{\rm max}^e=\frac{E^{e2}_F-m_e^{2}}{2 |q_e| B} \, , \label{eq:numaxe} \\
    \nu_{\rm max}^p=\frac{E^{p2}_F-M^{*2}}{2 q_p B} \, .  \label{eq:numaxp}
\end{align}

$k_{F,\nu}^q$ and $E_F^q$ are the Fermi momenta and energies of the particles, defined as 
\begin{align}
    k^{p}_{F,\nu}=&\sqrt{E^{p2}_F-M^{*2} - 2\nu q_p B} \, , \\
    k^{n}_F=&\sqrt{E^{n2}_F-M^{*2}} \, , \\
    k^{e}_{F,\nu}=&\sqrt{E^{e2}_F-m_e^{2}- 2\nu |e| B} \,.
\end{align}

The reader should note that the energy, defined above, and other thermodynamic quantities do not depend on $x_0$, the coordinate space defined as $x_0= p_y/m\omega_0$  with $\omega_0=qB/m$  the cyclotron frequency and $p_y$ the quantum number that specifies the $x$-projection of the guiding
center of the particle rotation, even though the hamiltonian of the system corresponds to a shifted harmonic oscillator by $x_0$.

The mean-field evaluation of the fields allows a closed expression for the bulk free energy density as:
\begin{equation}
    \mathcal{E}=\mathcal{E}_f+\mathcal{E}_p+\mathcal{E}_n \, ,
\end{equation}
where 
\begin{align} 
    \mathcal{E}_f=&\frac{m_\omega^2}{2}V_0^2 +\frac{\xi g_v^4}{8}V_0^4  + \frac{m_\rho^2}{2}b_0^2+\frac{m_\sigma^2}{2}\phi_0^2+\frac{\kappa}{6}\phi_0^3 \notag\\
    &+\frac{\lambda}{24}\phi_0^4+ 3 \lambda_{\omega\rho}g_\rho^2 g_\omega^2 V_0^2 b_0^2 \, , 
\end{align}
\begin{align}
    \mathcal{E}_n=&\frac{1}{4\pi^2}\left[k_{F}^nE_F^{n3}-\frac{1}{2}M^*\bigg(M^*k_{F}^nE_F^{n} \right.\notag \\
    &+M^{*3} \ln\bigg|\frac{k_{F}^n+E_F^{n}}{M^*}\bigg|\bigg)\bigg] \, , 
\end{align}    

\begin{align}
    \mathcal{E}_p=&\frac{q_pB}{4\pi^2}\sum_{\nu=0}^{\nu_{\rm max}}g_s\bigg[k_{F,\nu}^pE_F^{p}+\bigg(M^{*2}+2\nu q_pB \bigg) \notag \\
    &\cdot \ln\bigg|\frac{k_{F,\nu}^p+E_F^{p}}{\sqrt{M^{*2}+2\nu q_pB}}\bigg|\bigg] \, . \label{eq:enp}
\end{align}

Finally, the chemical potentials for protons, neutrons, and electrons are given by 
\begin{align}
    \mu_p=&E^p_F+g_\omega V^0+\frac{1}{2}g_\rho b^0 \, , \\
    \mu_n=&E^n_F+g_\omega V^0-\frac{1}{2}g_\rho b^0 \, , \\
    \mu_e=&E_F^e=\sqrt{k^{e2}_{F,\nu}+m_e^2+2\nu|q_e|B} \, .
\end{align}
and the baryonic pressure can be deduced as
\begin{equation}
   P=\mu_p\rho_p+\mu_n\rho_n-\mathcal{E}.
\end{equation}

 When considering charge-neutral, $\beta-$equilibrium matter, the following conditions should also be imposed:
\begin{eqnarray}
    \rho_p&=&\rho_e \, , \\
    \mu_n&=&\mu_p+\mu_e \, .
\end{eqnarray}

\subsection{Cluster and pasta structures in the CLD approximation}

In this work, we consider the CLD model \cite{pais15} to calculate the inner crust structures in $\beta-$equilibrium magnetized matter. We compare our results with a simpler CP calculation, that was also previously done in Ref.~\cite{wang22}. The reader should refer to this publication for further details on this calculation.

In the CLD model, each Wigner-Seitz cell 
is composed of a high-density ("cluster") part, labeled $I$, and a low-density ("gas") part, labeled $II$.  The equilibrium proportion of cluster and gas is obtained by minimizing the total energy density, including the interface surface and Coulomb terms, that is given by
\begin{equation}
\mathcal{E}=f\mathcal{E}^I+(1-f)\mathcal{E}^{II}+\mathcal{E}_{Coul}+\mathcal{E}_{surf}+\mathcal{E}_e \, ,
\label{En_dens}
\end{equation}
with respect to four variables: the linear size of the cluster, $R_d$, the baryonic density of the liquid phase $\rho^I$, the proton density of the liquid phase $\rho_p^I$, and the volume fraction of the liquid phase $f$. 
{The Coulomb and surface terms are given by 
\begin{eqnarray}
\mathcal{E}_{Coul}&=&2\alpha e^2\pi\Phi R_d^2 \left(\rho_p^I-\rho_p^{II}\right)^2 \, , \label{eq:ecoul} \\
\mathcal{E}_{surf}&=&\frac{\sigma\alpha D}{R_d} \label{eq:esurf}
\end{eqnarray}
where $\alpha=f$ for droplets, rods and slabs and $\alpha=1-f$ for tubes and bubbles. $\Phi$ is given by
\begin{eqnarray}
\Phi&=&\left(\frac{2-D\alpha^{1-2/D}}{D-2}+\alpha\right)\frac{1}{D+2} \, , \qquad D=1,3 \, ,  \nonumber \\
\Phi&=&\frac{\alpha-1-\ln\alpha}{D+2} \, , \qquad  D=2 \, .
\end{eqnarray}

The surface tension  parameter $\sigma$ depends on the total proton fraction of the system and its expression, for both EoS models used in this work, can be found in Ref.~\cite{Avancini12}. This parameter was obtained from a fit to a relativistic Thomas-Fermi calculation. For more details, the reader should check Ref.~\cite{Avancini12} and references therein.
}
{When we minimize $\mathcal{E}_{Coul}+\mathcal{E}_{surf}$ with respect to the size of the cluster, $R_d$, we get
\begin{eqnarray}
\mathcal{E}_{surf}&=&2\mathcal{E}_{Coul} \, ,   \\
R_d&=&\left[\frac{\sigma D}{4\pi e^2\Phi\left(\rho_p^I-\rho_p^{II}\right)^2 }\right]^{1/3} \, .
\end{eqnarray}

}

The {remaining} equilibrium conditions are

\begin{align}
    &\mu_n^I=\mu_n^{II} \, , \\
    &\mu_p^I=\mu_p^{II}-\frac{\mathcal{E}_{surf}}{(1-f)f(\rho^I_p-\rho^{II}_p)} \, , \\
    &P^I=P^{II}+\mathcal{E}_{surf}\bigg[\frac{3}{2\alpha}\frac{\partial\alpha}{\partial f}+\frac{1}{2\Phi}\frac{\partial\Phi}{\partial f}-\frac{((1-f)\rho_p^I+f\rho_p^{II})}{(1-f)f(\rho^I_p-\rho^{II}_p)}\bigg] \, .
\end{align}
The reader can check e.g. Ref.~\cite{wang22} for the different expressions in the total energy density. 
We note that we do not consider superfluidity, since it has already been shown in previous studies \cite{Burrello15} that its impact on the static properties of a neutron star is very small.

\begin{figure*}
	\begin{center}
		\begin{tabular}{cc}
			\includegraphics[width=0.45\linewidth,angle=0]{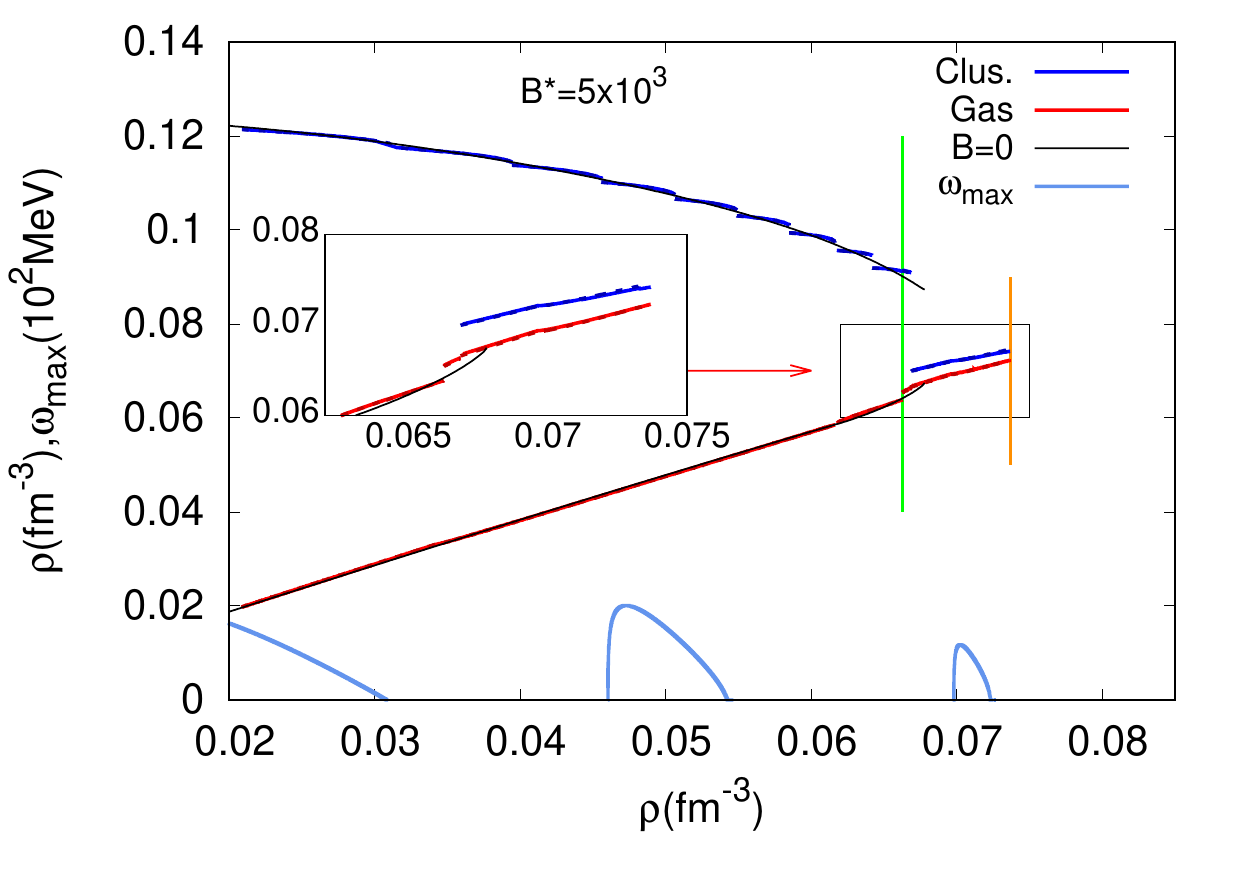}&
			\includegraphics[width=0.45\linewidth,angle=0]{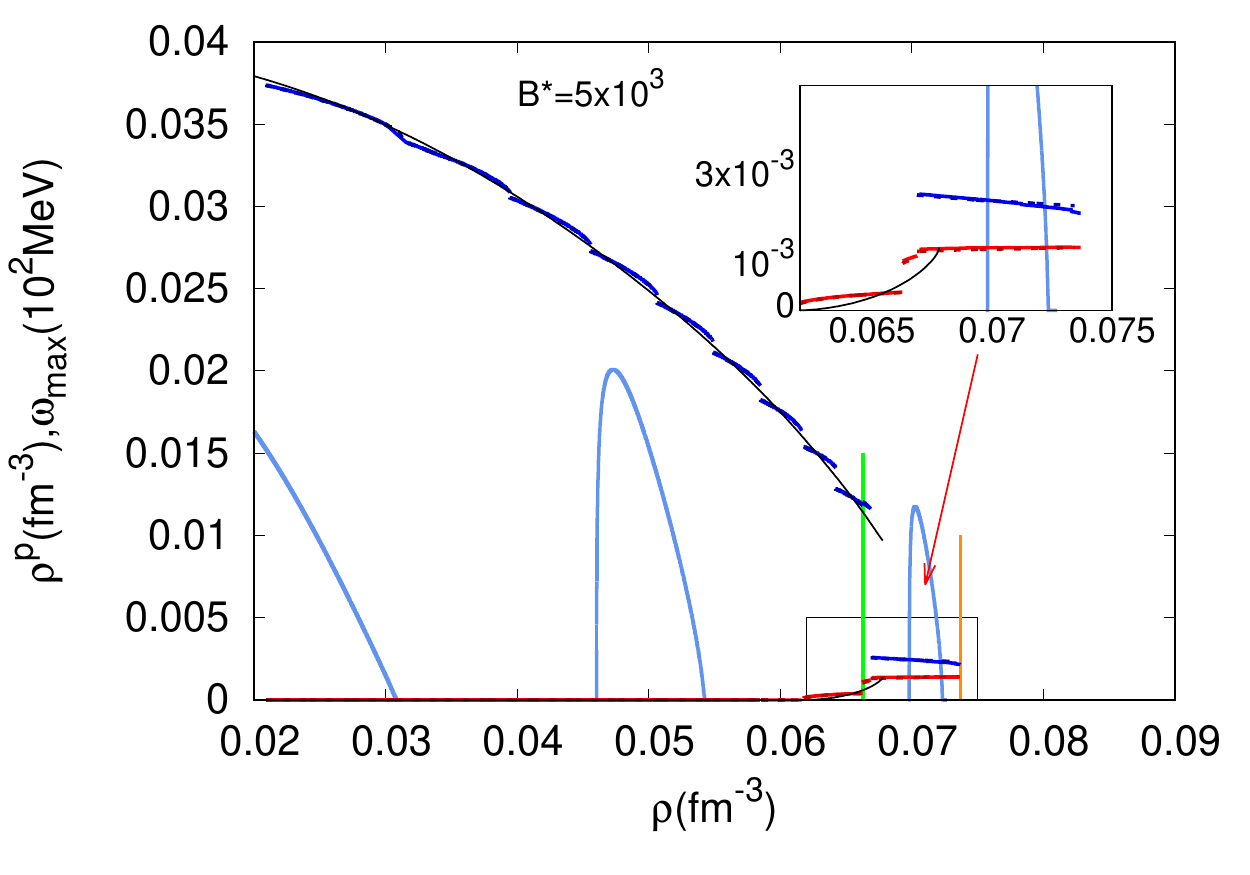}\\
			\includegraphics[width=0.45\linewidth,angle=0]{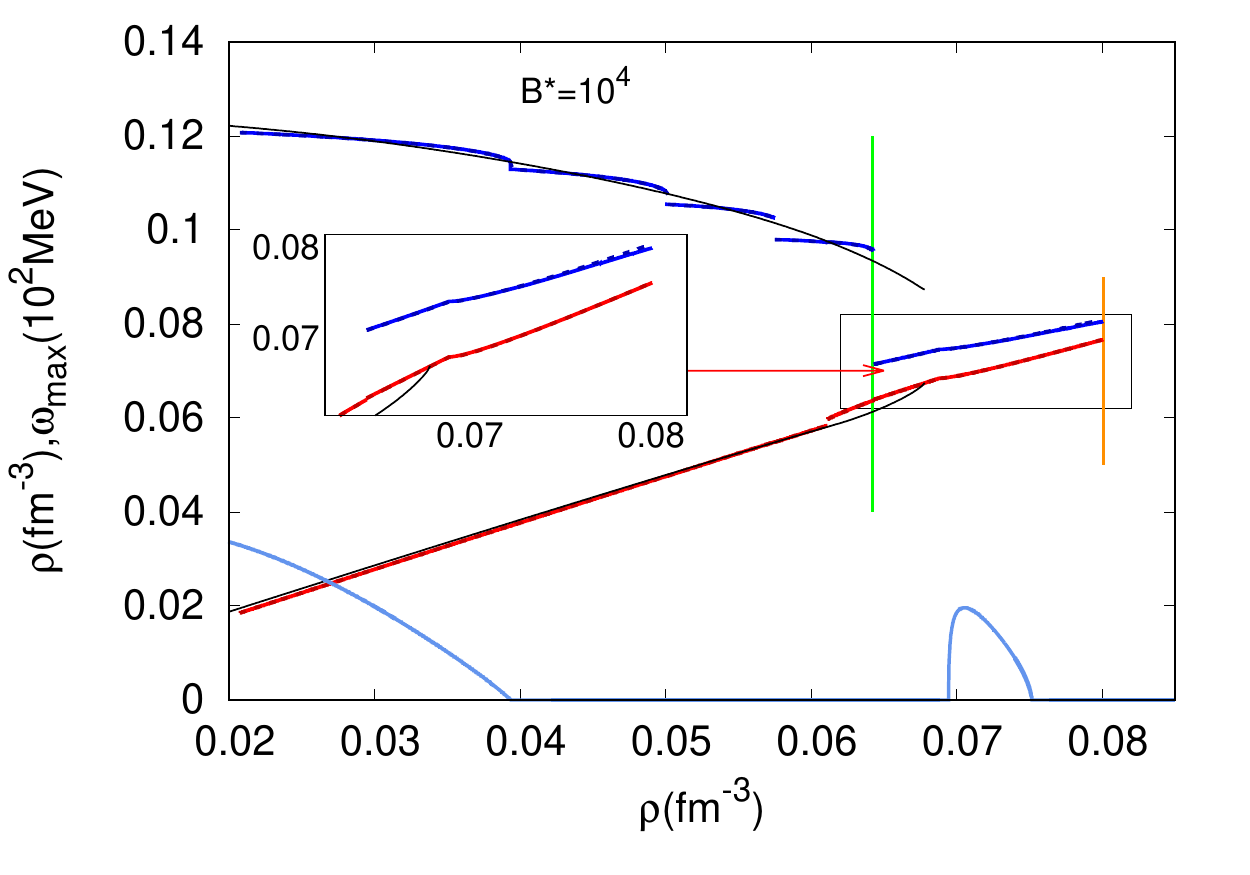} &
			\includegraphics[width=0.45\linewidth,angle=0]{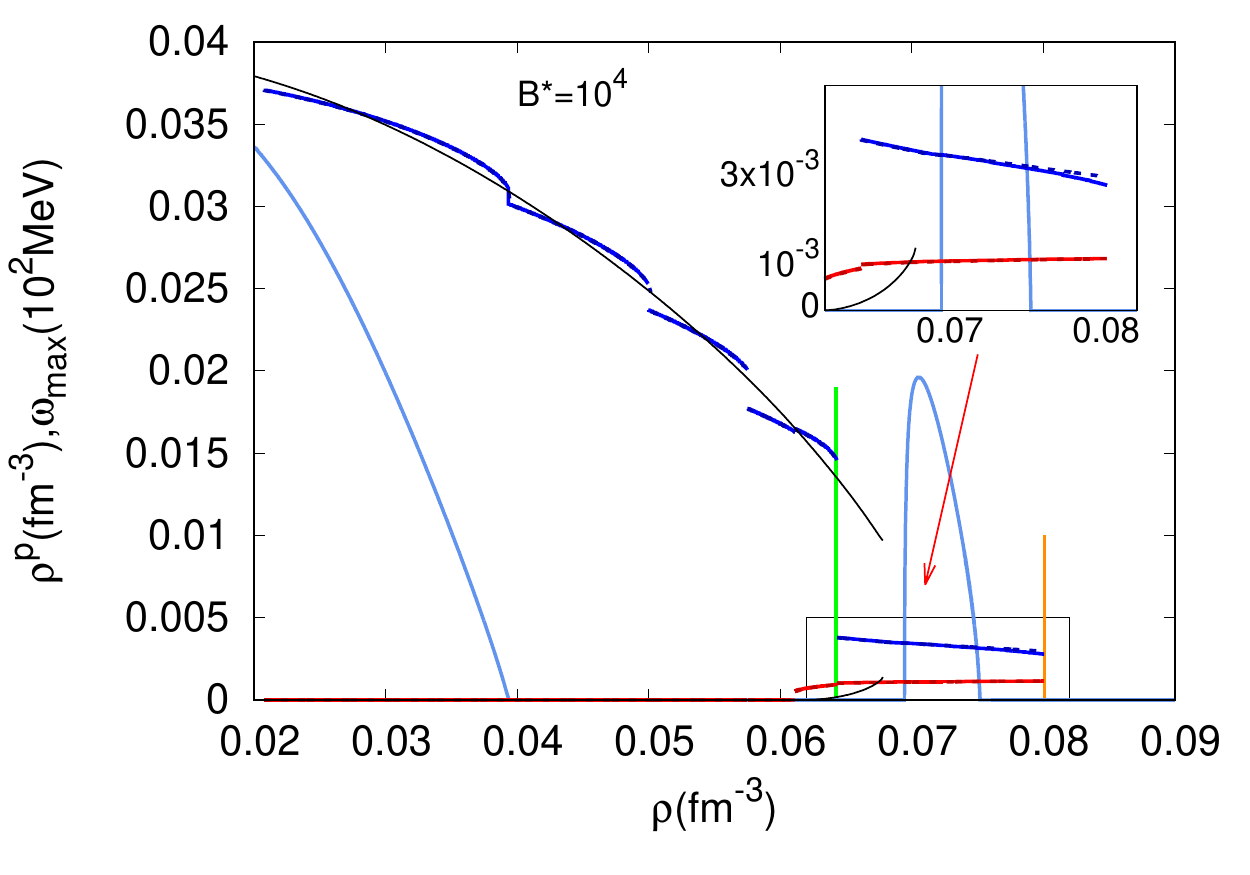}
		\end{tabular}
	\end{center}
	\caption{Baryon (left) and proton (right) densities of liquid (1, blue) and gas (2, red) phases as function of the total baryon density for the NL3 model in a CP (dashed) and CLD (solid) calculations, with $B^*=5\times 10^3$ (top) and $B^*=10^4$ (bottom). We also plot the magnetized growth rates divided by a factor $10^2$, $|\omega_{\rm max}|$ (light blue), as well as the densities in the $B=0$ case (black). The green and orange segments indicate respectively $\rho_{1\rightarrow 2}$ and $\rho_{cc}$, both defined in the text.}
	\label{fig2}
 \end{figure*}

\section{Results and discussion}

\begin{figure*}[htbp]\centering
	\begin{center}
		\begin{tabular}{cc}
			\includegraphics[width=0.45\linewidth,angle=0]{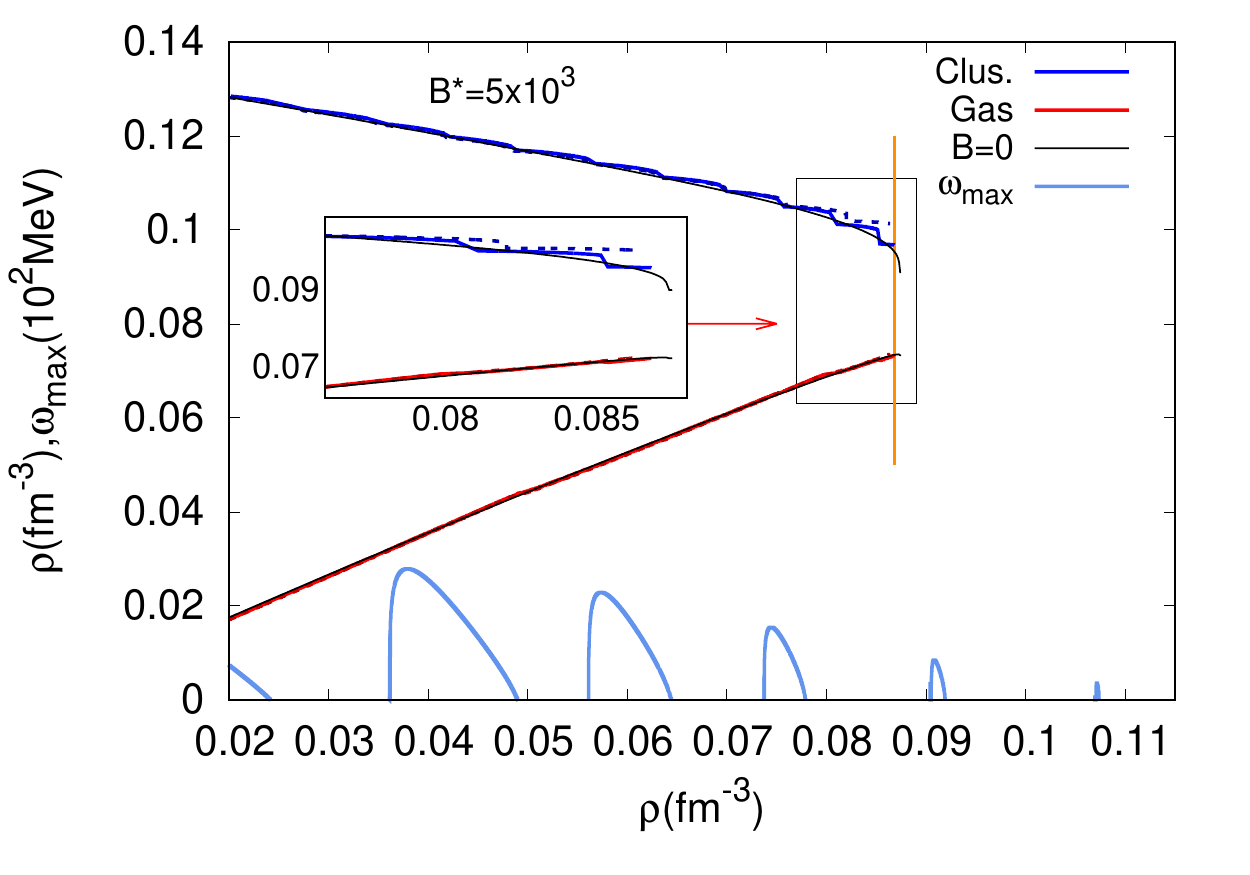}&
			\includegraphics[width=0.45\linewidth,angle=0]{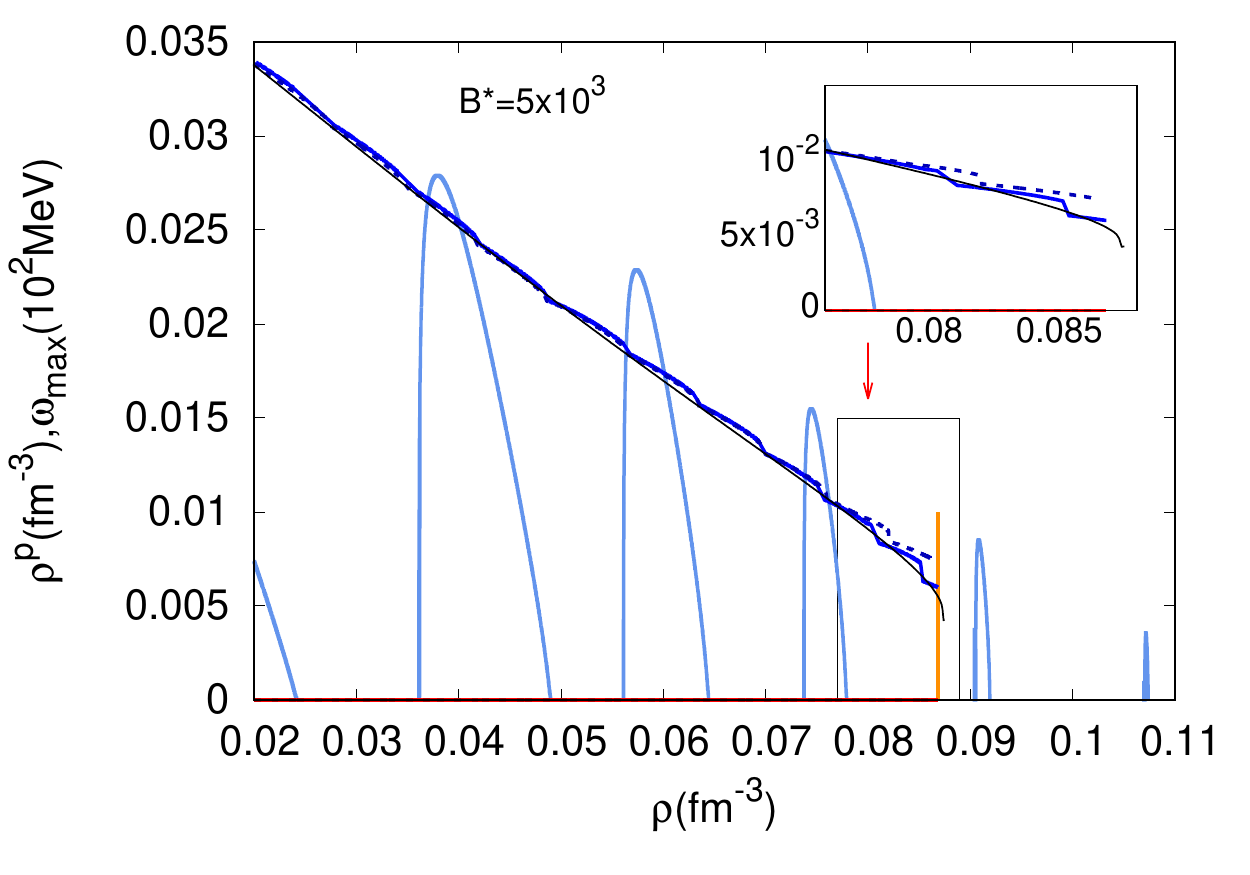}\\
			\includegraphics[width=0.45\linewidth,angle=0]{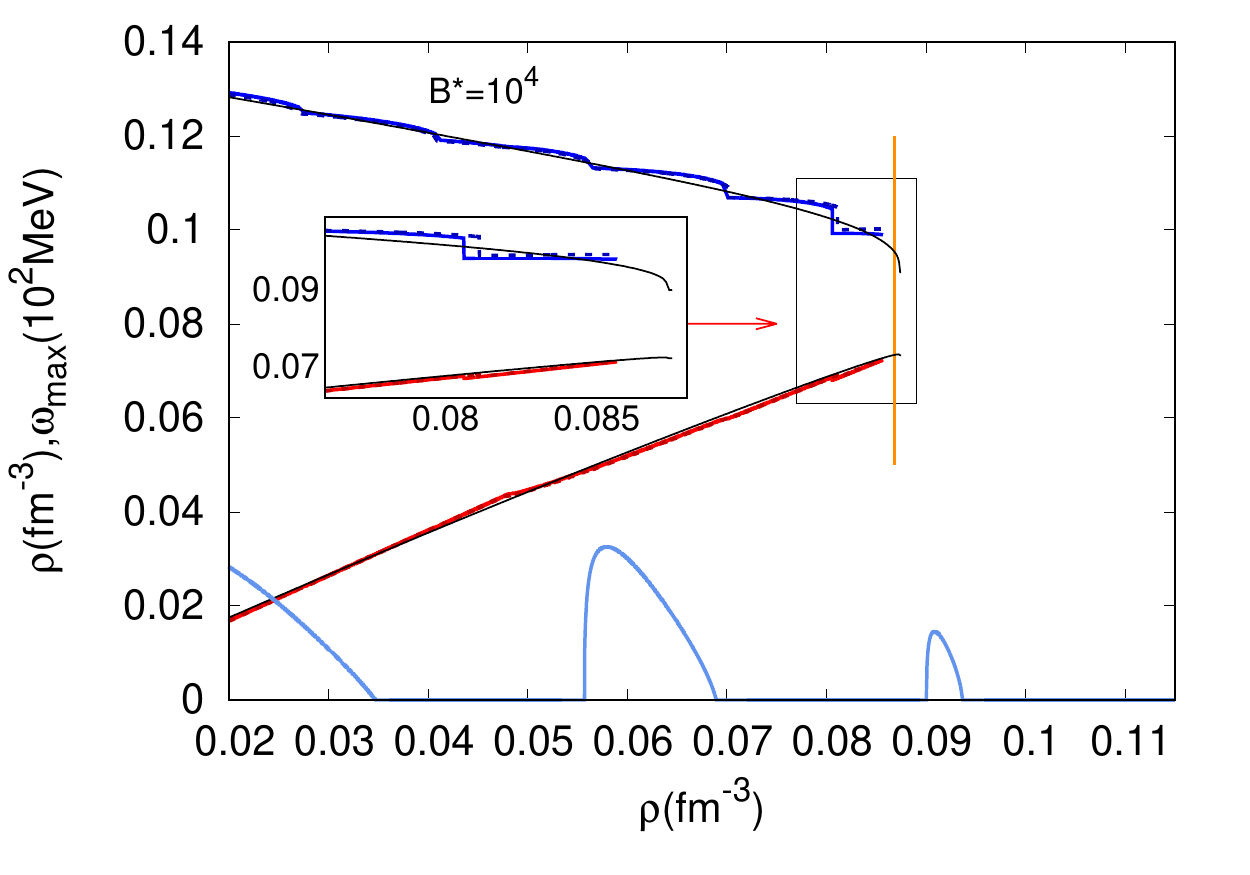} &
			\includegraphics[width=0.45\linewidth,angle=0]{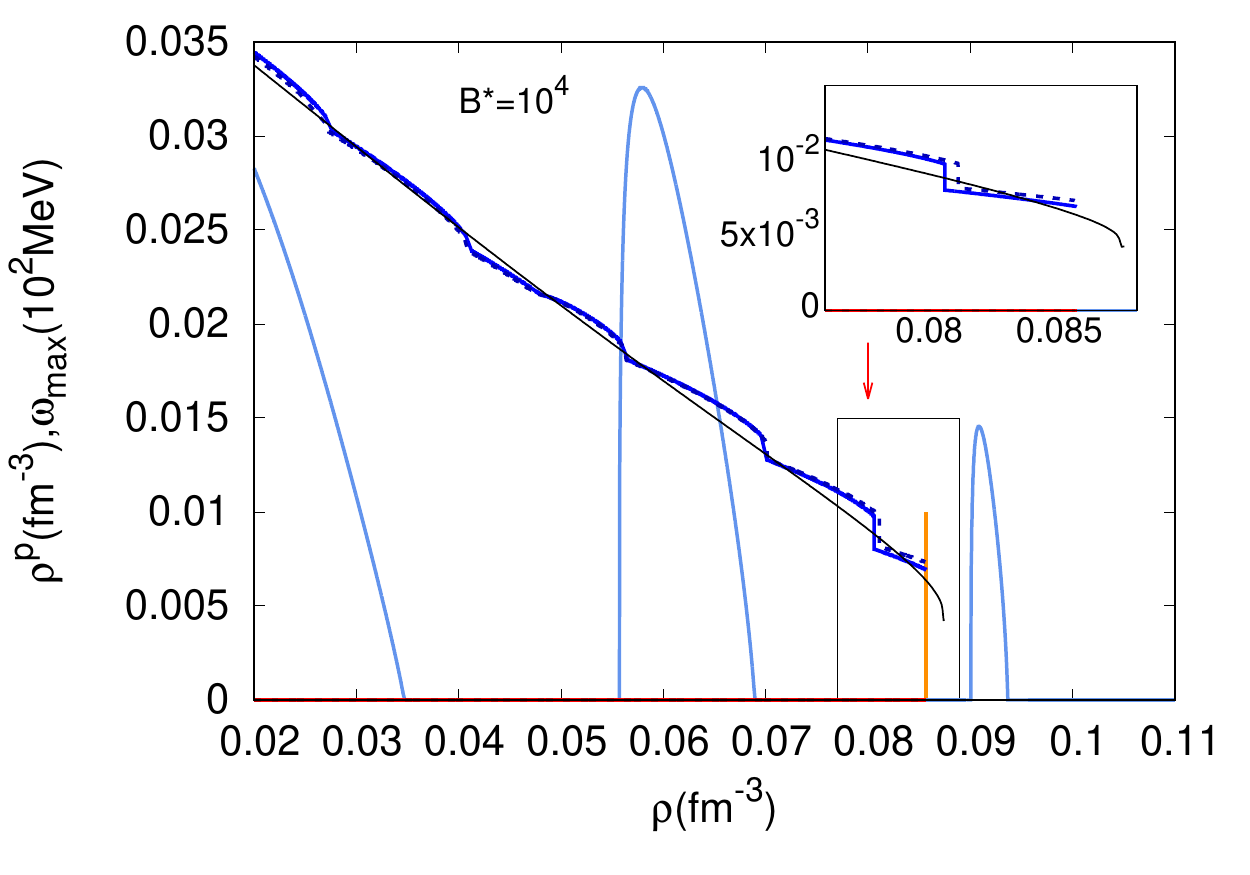}
		\end{tabular}
	\end{center}
	\caption{Baryon (left) and proton (right) densities of liquid (1, blue) and gas (2, red) phases as function of the total baryon density for the NL3$\omega\rho$ model in a CP (dashed) and CLD (solid) calculations, with $B^*=5\times 10^3$ (top) and $B^*=10^4$ (bottom). We also plot the magnetized growth rates divided by a factor $10^2$, $|\omega_{\rm max}|$ (light blue), as well as the densities in the $B=0$ case (black). The orange segments indicate $\rho_{cc}$, defined in the text.} 
	\label{fig3}
 \end{figure*}

 In this section, we analyze the results obtained in our study, comparing them to previous calculations, in particular, the analysis of the dynamical spinodal in Refs.~\cite{fang16,fang17}, and the study done with the CP model in \cite{wang22}. We will show how the CLD model, while in agreement with the previous calculations, allows a more realistic description of the system, and helps clarifying some contradictory results that can be found in the literature~\cite{bao}. 
 
 We consider stellar matter in $\beta-$equilibrium, taking the NL3 and the NL3$\omega\rho$ RMF models, for different values of the magnetic field. We will consider a magnetic field of $B^*=5\times 10^3$ and $B^*=10^4$, corresponding respectively to 
 $B=2.2\times 10^{17}$ G and $B=4.4\times 10^{17}$ G. These values are the same as in Ref.~\cite{wang22}.

In Fig.~\ref{fig1}, we plot the difference between the energy per baryon of homogeneous matter and the one of clustered matter, 

comparing the results obtained with the CP and the CLD calculations. The results for both values of the magnetic field are shown. The density at which this difference crosses zero indicates the transition between homogeneous (core-like) and non-homogeneous (crust-like) matter. From this Figure we can extract the following conclusions: the two calculations give a similar value of the crust-core transition density; while the CP calculation (left panels) tends to give both stable solutions, i.e. solutions in which the clusterized matter energy is lower than the one of homogeneous matter, and metastable solutions, i.e. solutions in which the energy is higher than the one of homogeneous matter, in the CLD calculation (right panels), the metastable solutions are almost completely absent.
This feature is particularly pronounced at the highest value of the field (lower panels) and for the softest EoS (dashed lines). This finding suggests that the estimation of the transition density might be overestimated, if the CP approach is considered.

In Fig.~\ref{fig2}  we show the baryon (left) and proton (right) densities of the clusterized (blue) and non-clusterized (red) part of the Wigner-Seitz cell, together with the growth rates $\omega_{\rm max}$ obtained from the study of the dynamical spinodal, for both values of the magnetic field, considering the NL3 model. The growth rates are taken from \cite{fang17}, where the classical Vlasov approach was used. 
They are included in the plot in order to compare the new results to the ones in literature.
{In this figure, the difference between the results given by the CP and CLD calculations is always negligible compared to the scale of the plot, making the two lines almost indistinguishable.} 

It can be seen from the plots that, as already showed in Ref.~\cite{wang22}, in the presence of the magnetic field, the inner crust can be divided into two regions. In the first region, the density inhomogeneity is important, even if it decreases with the baryon density, and the protons are confined inside the cluster. This is the standard feature of a solid crust, corresponding to a periodic arrangement of the ions in the lattice, and consequently of the electrostatic potential.  

A second, extended region, only appears in the finite-$B$ calculation, and starts approximately at the $B=0$ crust-core transition density. 
In this region, the inhomogeneous solution is still energetically favored over the homogeneous one, but the amplitude of the inhomogeneity is very small. 

Indeed, the density fluctuation $(\rho_{clus}-\rho_{gas})/(\rho_{clus}+\rho_{gas})$ is in the range $1.3-5.5\%$ and the proportion of dripped protons is significant with respect to the fraction of protons bound in the cluster, with proton density fluctuations $(\rho^p_{clus}-\rho^p_{gas})/(\rho^p_{clus}+\rho^p_{gas})$ in the range $21-57\%$.
We can also see that at the border between the crust and this region, the proton and baryon cluster densities discontinuously jump.  
The equilibrium composition of matter obtained through the CLD variational calculation is nicely reflected in the behavior of the dynamical spinodal (curves labelled ``$\omega_{\rm max}$" in Figs.~\ref{fig2} \ref{fig3}), which suggests that the classical Vlasov approach is able to describe, at least at a qualitative level, the behaviour of the system. A first, bigger region of instability can be seen in the region corresponding to the $B=0$ calculation, where the densities are very different, and a second, smaller region of instability, with a smaller value of the growth rate, which appears only in the finite-$B$ calculation, can be seen after the discontinuous jump of the cluster density.
If the crust-core transition is defined as the transition point between the solid and the liquid part of the star, these findings suggest that the ``extended crust" deduced from the spinodal analysis \cite{rabhi2,fang16,fang17,chen17} might be rather interpreted as a peculiar inhomogeneous liquid portion of the outer core.
This shows the limits of the Vlasov approach, that neglects both quantum and non-linear effects.
{In the following we will refer to the transition density between the first and the second region as $\rho_{1\rightarrow 2}$, marked in the figure by a green segment, while we will refer to the transition density between inhomogeneous and homogeneous matter as $\rho_{cc}$, marked in the figure by an orange segment. The values for both densities in the various cases can be found in Tab.~\ref{Tab2}. }

The analysis presented in Fig.~\ref{fig2} is repeated in Fig.~\ref{fig3} in the case of the NL3$\omega\rho$ model. Also in this case, the CLD and CP results are indistinguishable on the scale of the figure as in Fig.~\ref{fig2} above. For the NL3$\omega\rho$ model, 
we see that the extended region does not appear and $\rho_{1\to 2}\equiv\rho_{cc}$.
A tendency towards a liquid region with small amplitude inhomogeneities can still be seen as a (small) density drop correlated to the instability observed in the spinodal analysis, but apparently the homogeneous matter instability is not sufficient to produce an equilibrium inhomogeneity.
These results are in agreement with the ones found before \cite{wang22}, but in Wang et al. this energy criterium was not used, and this explains why the authors interpreted the results as the persistence of the inhomogeneous region to higher densities as compared to this work.

{In Tab.~\ref{Tab2}, we present the values for $\rho_{1\rightarrow 2}$ and $\rho_{cc}$ for the different values of the magnetic field and for the two models.}

In the NL3 model, the higher the value of $B$, the higher the value of the transition density to homogeneous matter $\rho_{cc}$, and in
the NL3$\omega\rho$ model, the opposite happens, i.e. the higher
the value of $B$, the smaller the value of the 
transition density.  In fact, this effect is due to the extended region, that only appears in the NL3 model. If we  discard this region, and consider the crust-core transition as the the opening of proton drip 
$\rho_{1\rightarrow 2}$, this density decreases with increasing $B$, as in the NL3$\omega\rho$ model. 

\begin{figure}[htbp]\centering
	\begin{center}
		\begin{tabular}{c}
			\includegraphics[width=\linewidth,angle=0]{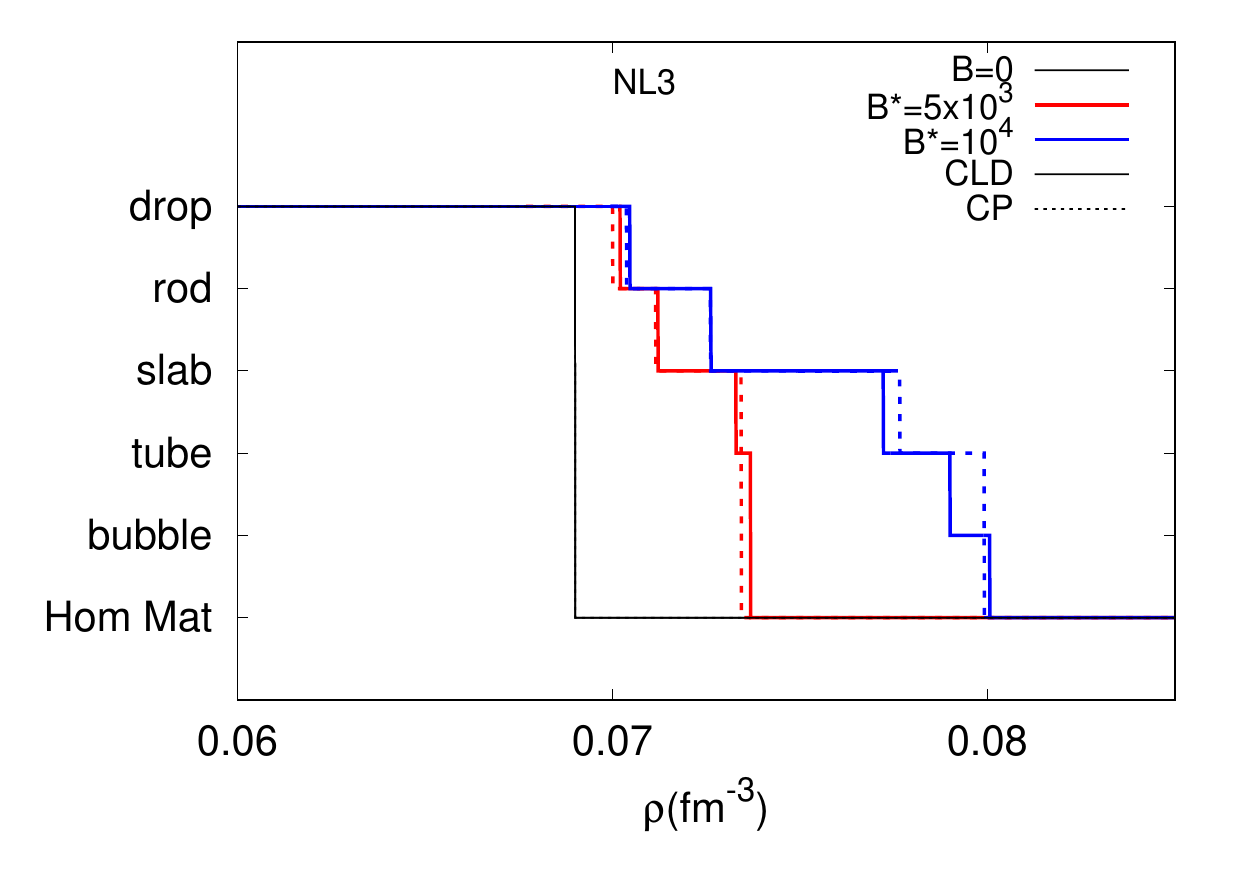}\\
			\includegraphics[width=\linewidth,angle=0]{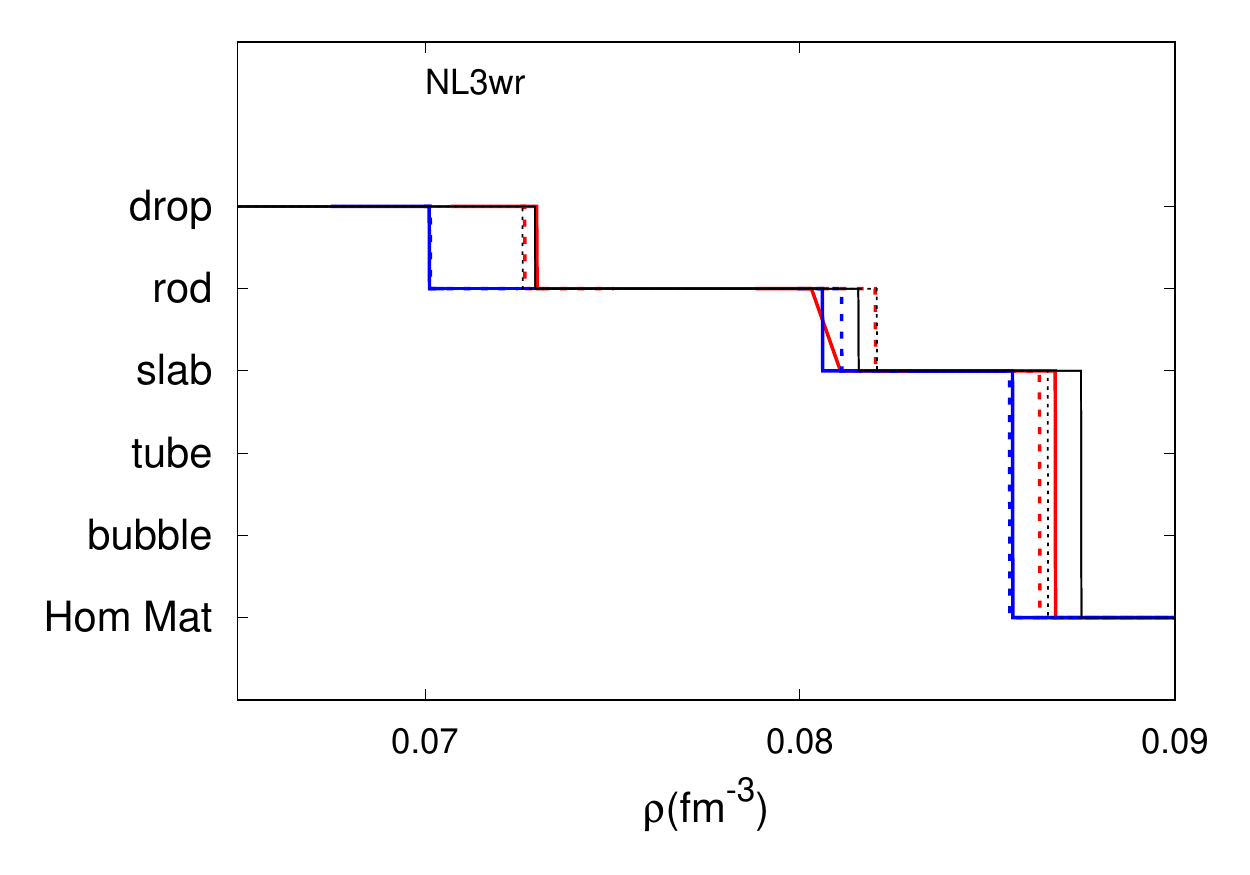}
		\end{tabular}
	\end{center}
	\caption{Geometry of the clusters as a function of the total baryon density for the NL3 (top) and NL3$\omega\rho$ (bottom) models. The results for $B=0$ (black), $B^*=5\times 10^3$ (red) and for $B^*=10^4$ (blue) are displayed, as well as the results for the CP (dashed line) and CLD (solid line) calculations.}
	\label{fig5}
 \end{figure}

The different behavior of the two models with respect to the magnetic field can be also appreciated by looking at Fig.~\ref{fig5}, where the geometry of the clusters is plotted as a function of the baryon density. We observe that, for all the calculations considered, the CLD model gives  slightly higher transition densities between the different shapes, as well as to homogeneous matter ( values displayed in Tab.~II as $\rho_{cc}$).

\begin{table*}[]
    \centering
    \begin{tabular}{c|c|cccc}
        \hline
        \hline
         \multicolumn{2}{c}{}&  \multicolumn{2}{c}{$\rho_{1\rightarrow2}$}& \multicolumn{2}{c}{$\rho_{cc}$} \\
         \multicolumn{2}{c}{} & CP& CLD &CP & CLD\\
         \hline
         \multirow{3}{1.5cm}{NL3}& B=0 & - & - & 0.06865 & 0.06780  \\
          & $B^*=5\times 10^3$ & 0.06627 & 0.06627 & 0.07344 & 0.07369\\
          & $B^*=10^4$ & 0.06431 & 0.06421 & 0.07992 & 0.08007 \\
          \hline
          \multirow{3}{1.5cm}{NL3$\omega\rho$}& B=0 & - & - & 0.08662 & 0.08751 \\
          & $B^*=5\times 10^3$ & - & - & 0.08640 & 0.08682 \\
          & $B^*=10^4$ & - & - & 0.08560 & 0.08568\\
          \hline
          \hline
    \end{tabular}
    \caption{Crust-core transition densities (cc) and densities for the transition from the solid (1) to the liquid (2) region of the inner crust, in units of fm$^{-3}$, for $\beta-$equilibrium matter, considering the NL3 and the NL3$\omega\rho$ models in the CP and CLD approximations, and taking $B=0$, $B^*=5\times 10^3$ and $B^*=10^4$. }
   \label{Tab2}
\end{table*}

\begin{figure}[htbp]\centering
	\begin{center}
		\begin{tabular}{c}
			\includegraphics[width=0.95\linewidth,angle=0]{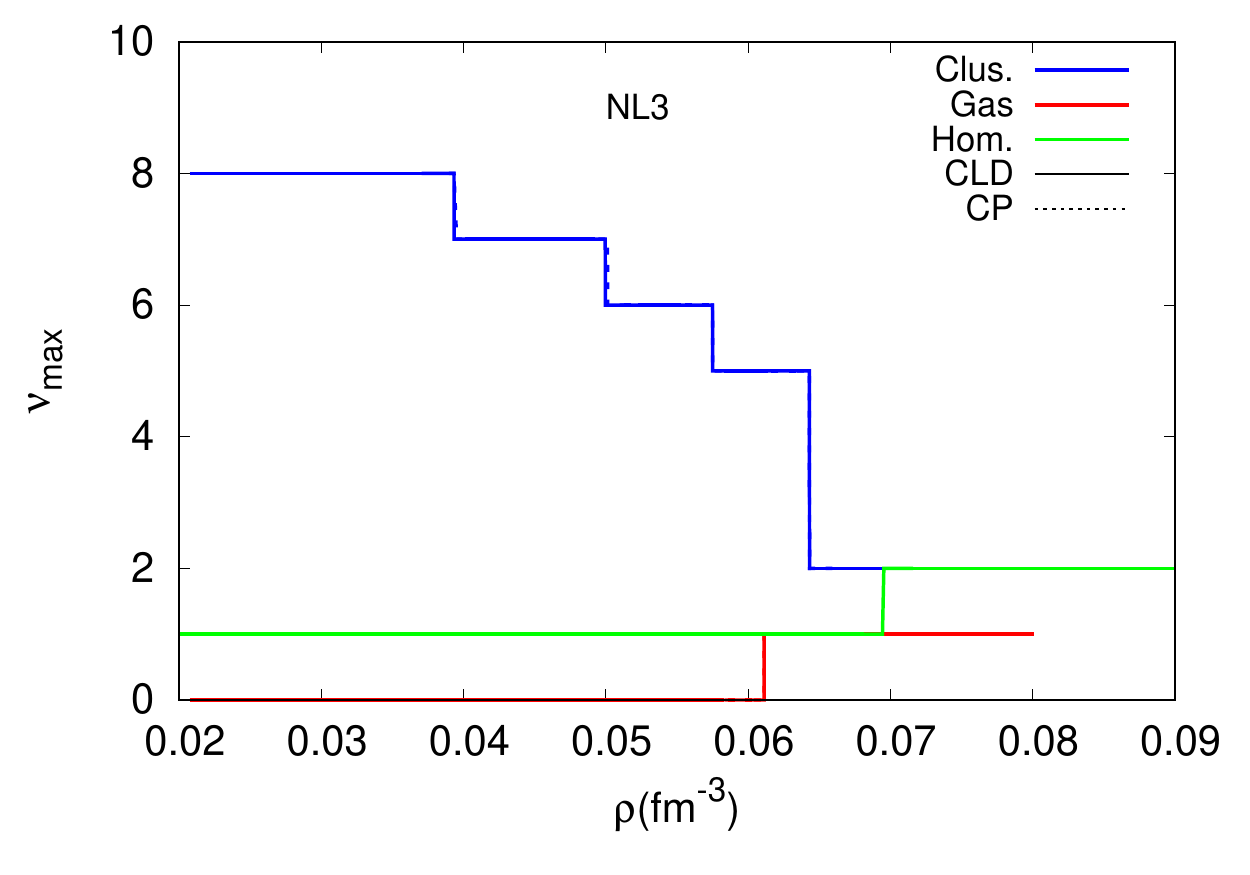} \\
			\includegraphics[width=0.95\linewidth,angle=0]{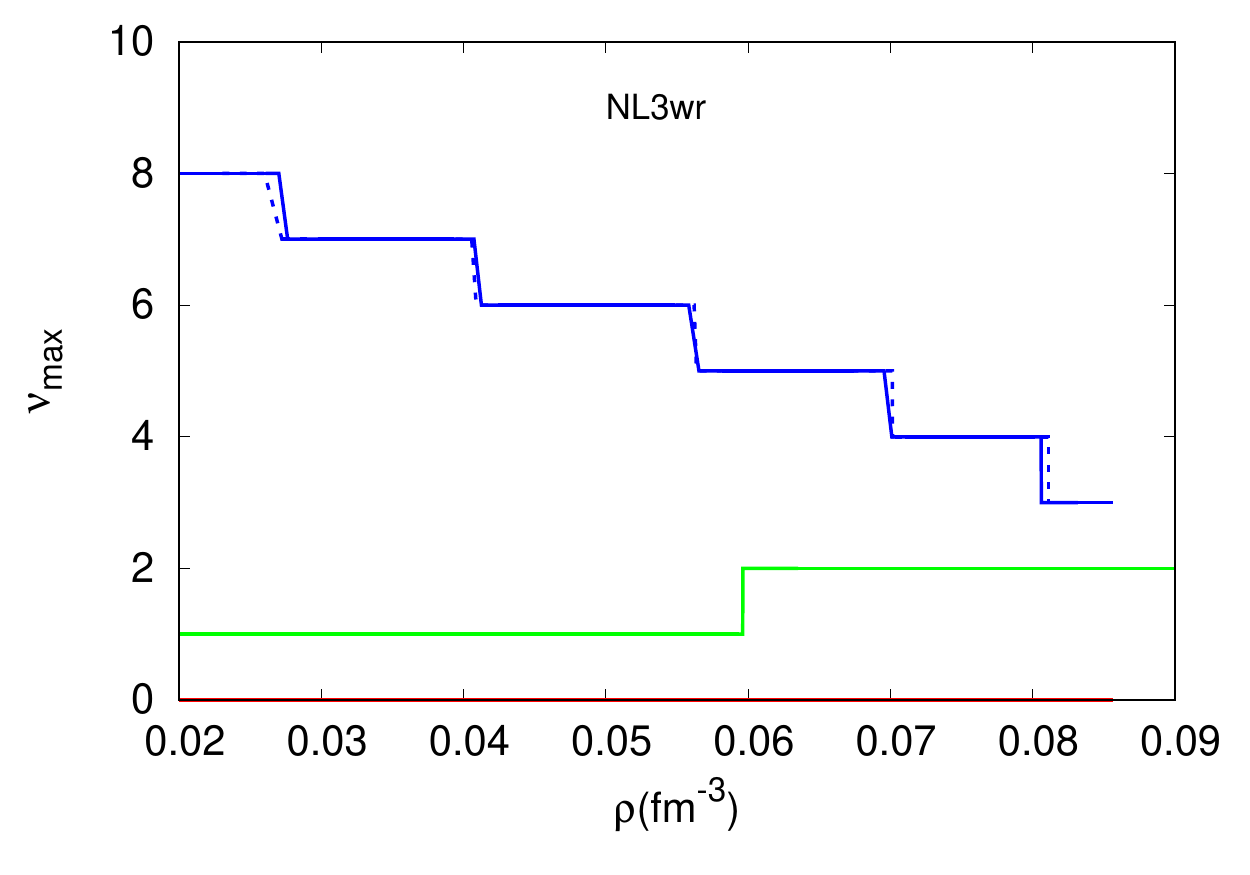}
		\end{tabular}
	\end{center}
	\caption{Proton $\nu_{\rm max}$ of high (blue, marked as "Clus") and low (red, marked as "Gas") density phases, together with the proton $\nu_{\rm max}$ of homogeneous matter (green, marked as "Hom."), as function of the total baryon density for the NL3 (top) and NL3$\omega\rho$ (bottom) models in a CP (dashed line) and CLD (solid line) calculations, with $B^*=10^4$. }
	\label{fig4}
 \end{figure}

{The fact} that small discontinuities appear in the density behavior of the clustered region at the highest densities, particularly for the higher value of the magnetic field (see inserts in Fig.~\ref{fig3})
suggests that the physical origin of the extended region might be related to the discontinuous occupation of the Landau levels. 
To better understand this peculiar thermodynamic behavior, where a spinodal instability does not lead to phase separation but to an equilibrium configuration with small amplitude inhomogeneities,  we plot in Fig.~\ref{fig4} the density evolution of the number of Landau levels (LL), in the highest $B$ case. 
 The other value of $B$ gives the same qualitative information. 
Both models are considered, such as to understand the origin
of the model dependence of the results.

In that Figure, the maximum number $\nu_{\rm max}$  of occupied LL for the protons Eq.~(\ref{eq:numaxp}) is displayed. This is calculated for the higher density region I of Eq.(\ref{En_dens}) (blue curves labelled "Clus" in Figure \ref{fig4}), the lower density region II (red curves labelled "Gas"), and also for homogeneous nuclear matter (green curves labelled "Hom."). The CLD calculations are given by solid curves, while CP ones (almost indistinguishable from CLD) correspond to dashed curves.

As a general statement, the number of occupied LL for a particle $q=e,p$
increases with its density, and so does the kinetic energy of the system. 
Because of charge neutrality, the behavior of $\nu_{\rm max}^e$ is the same as the behavior of $\nu_{\rm max}^p$ in the case of an equivalent perfectly homogeneous system. In the case of NL3 (top panel) we can see that  preserving the inhomogeneity in the extended region does not increase  the   number of occupied LL in the denser part with respect to homogeneous matter (because the amplitude of the inhomogeneity is very small), and yet is enough to allow the protons in the dilute phase to stay within the first LL. As a consequence, the effective Fermi energy is reduced with respect to the homogeneous configuration, and this energy gain is enough to compensate the Coulomb and surface energy cost associated to the density fluctuation.
In the case of NL3$\omega\rho$, the lower value of $L$ implies a higher symmetry energy with respect to NL3. In turn, this leads to a higher global proton fraction,  as already pointed out in Ref.~\cite{wang22}, and shown by the fact that the second proton  LL is already occupied at relatively low densities, and the third LL is also filled in the denser phase.    As a consequence, keeping the density fluctuation beyond the crust-core transition point of the $B=0$ case implies a higher cost in terms of effective Fermi energy, and the instability of homogeneous matter towards density fluctuations observed in the spinodal analysis is effectively damped.

\begin{table*}  
    \begin{tabular}{c|c|c|ccccc}
    \hline
    \hline
          \multicolumn{3}{c}{}&$M_1(M_\odot)$ & $M_2(M_\odot)$ &  $R_T(km)$ & $\Delta R_1(km)$ & $\Delta R_2(km)$ \\
     \hline
         \multirow{6}{1.5cm}{$M_T=1.4 M_\odot$}& \multirow{3}{1.5cm}{NL3} & $B=0$ & 0.0588 & 0.0 & 14.685 & 1.4270 & 0.0  \\
          &&$B^*=5\times10^3$ &  0.0597 & 0.0258 & 14.908 & 1.6532 & 0.1541\\
          &&$B^*=10^4$ &  0.0574 & 0.0414 & 15.025 & 1.7427 & 0.3148\\
         \cline{2-8}
          &\multirow{3}{1.5cm}{NL3$\omega\rho$} & B=0 & 0.0457 & 0.0 & 13.747 & 1.3665 & 0.0 \\
          &&$B^*=5\times 10^3$ &  0.0526 & 0.0 & 13.871 & 1.5431 & 0.0\\
          &&$B^*=10^4$ &  0.0526 & 0.0 & 13.991 & 1.6556 & 0.0\\
         \hline
        \multirow{6}{1.5cm}{$M_T=2.0 M_\odot$}& \multirow{3}{1.5cm}{NL3}& $B=0$ & 0.0394 & 0.0 & 14.777 & 0.8691 & 0.0  \\
         & & $B^*=5\times10^3$ &  0.0400 & 0.0132 & 14.914 & 1.0064 & 0.0932\\
         & & $B^*=10^4$ &  0.0385& 0.0288 & 14.989 & 1.0617 & 0.1973\\
         \cline{2-8}
         & \multirow{3}{1.5cm}{NL3$\omega\rho$}& B=0 & 0.0326 & 0.0 & 14.079 & 0.8632 & 0.0  \\
         & & $B^*=5\times 10^3$ &  0.0384 & 0.0 & 14.161 & 0.9769 & 0.0\\
         & & $B^*=10^4$ &  0.0383 & 0.0 & 14.234 & 1.0437 & 0.0\\
         \hline
         \hline
    \end{tabular}
    \caption{Mass and radial width of the solid (1) and liquid (2) crust of the NS, together with the total radius of the star $R_T$, for the NL3 and NL3$\omega\rho$ model in the CLD approximation, and taking $B=0$, $B^*=5\times 10^3$ and $B^*=10^4$, for two different values of the total mass of the star. See text for the definition of the different quantities. }   \label{Tab3}
\end{table*}

To conclude our analysis, we show the effects of the modification of the crust structure due to the magnetic field, on the static properties of neutron stars. To this aim,  we built a unified EoS, by calculating the EoS of homogeneous npe matter in the presence of magnetic field, using the same RMF models for the inner crust. Concerning the {magnetized} outer crust, we used the code by Chamel et. al \cite{Chamel_code}, which was already used to obtain the results in \cite{Chamel20}. {The outer core} is defined as usual as the region extending from the star radius inward until the neutron drip point; the ``solid" part of the inner crust is assumed to end at the point $\rho_{1\to 2}$ where the cluster density discontinuously drop; the ``liquid" part of the inner crust is assumed to coincide with the extra inhomogeneous region ending at the density point $\rho_{cc}$ where homogeneous matter becomes energetically favored.

{In Tab.~\ref{Tab3}, we report the spatial extension of the two different regions of the star, together with the associated masses and the total radius of the star. 

The results for two different values of the total mass of the star are shown. }
{We should note that for a completely consistent calculation, the coupled Maxwell-Einstein equations  should have been computed (see e.g. Ref.~\cite{gomes19}, where the authors used the LORENE library to integrate those equations). We also underline that here we assume force-free, constant and uniform magnetic field throughout the star, neglecting the effects of electric current and associated phenomena within the star.}
 Even if the absolute values of the radii reported in Tab.~\ref{Tab3} are not fully realistic, the qualitative behavior appears correct. 
In particular we observe, in agreement with Ref.~\cite{gomes19}, an increase of the crustal mass and radius with the magnetic field.

From the Table, it can be seen that 
in the case of NL3 and for high B, the mass of the ``liquid" crust becomes comparable to the one of the ``solid" part.

\section{Conclusions}	

In this paper, we studied the structure of the inner crust of a neutron star in the presence of a strong magnetic field, within a relativistic mean-field framework, and using the compressible liquid drop model for the calculation of the pasta phases. We then compared our results with the ones obtained in previous studies using the coexisting phases calculation \cite{pais21,wang22}, and the dynamical spinodal method \cite{fang16,fang17}. We considered two different RMF models, NL3 \cite{lalazissis97} and NL3$\omega\rho$ \cite{horowitz01}, and two different values of the magnetic field, namely $B=1.3\times 10^{17}$ G and $B=4.4\times 10^{17}$ G.

Our main result is that the extended spinodal instability observed in different previous works \cite{rabhi2,fang16,fang17,chen17,fang17a,pais21} leads to stable or metastable equilibrium configurations that are  inhomogeneous, with density fluctuations $(\rho_{clus}-\rho_{gas})/(\rho_{clus}+\rho_{gas})$ in the range $1.3-5.5\%$ and proton density fluctuations $(\rho^p_{clus}-\rho^p_{gas})/(\rho^p_{clus}+\rho^p_{gas})$ in the range $21-57\%$.
The energetic gain of such small amplitude fluctuations is due to the possibility, for the protons of the more dilute regions, of occupying a lower order LL. Because of that, the existence of such inhomogeneous configurations depends crucially on the proton fraction, and therefore 
on both the strength of the magnetic field, and on the value of the symmetry energy in the sub-saturation region. In particular, the configurations are stable only if the slope parameter $L$ is high, as in the case of the NL3 functional.

We found that the transition densities given by the CLD calculation 
are in good agreement with the simpler CP approximation employed in the previous analysis \cite{wang22}. 

The presence of metastable inhomogeneous solutions enlighted in the present work, particularly for the softer NL3$\omega\rho$ model, may explain why different results on a possible extended crust in magnetars were reported in the literature depending on the chosen functionals and crust modelling  \cite{rabhi2,fang16,fang17,chen17,fang17a,pais21,wang22,bao}.

Moreover, the qualitative effect of the magnetic field on the crust-core transition densities is also in good agreement with the Thomas-Fermi calculations by Bao et al.~\cite{bao} for the cases where the extended crust does not appear, such as the NL3$\omega\rho$ model. 

However, its contribution can amount to approximately $4\%$ of the mass and $6\%$ of the radius of a heavily magnetized canonical $1.4 M_\odot$ neutron star, for an EOS as stiff as NL3. If $L$ values as high as the ones proposed by recent analyses of PREX-2 data \cite{Adhikari21} were to be confirmed in the future, it will be very important to study the elasticity and conductivity properties of this intermediate region, in order to settle its possible influence on NS observations.

\section*{ACKNOWLEDGMENTS}
This work was partly supported by the FCT (Portugal) Projects No. 2022.06460.PTDC and UID/FIS/04564/2020, and  by the IN2P3 Master Project NewMAC. L.S. acknowledges the PhD grant 2021.08779.BD (FCT, Portugal), and LPC-Caen for support.
H.P. acknowledges her former grant CEECIND/03092/2017 (FCT, Portugal) while at CFisUC, Department of Physics, University of Coimbra, 3004-516 Coimbra, Portugal, where this work was mostly done.

\end{document}